\documentclass[preprint,12pt]{elsarticle}

\usepackage{amssymb}
\usepackage{lscape}
\usepackage{setspace}
\usepackage{graphicx}
\usepackage{amsmath,amsfonts,amssymb}

\journal{}

\begin{document}
\begin{spacing}{2.0}
\begin{frontmatter}



\title{$Ab$ $intito$ study on some new spin-gapless semiconductors: The Zr-based quanternary Heusler alloys
}


\author[mymainaddress]{Qiang Gao}
\author[mymainaddress]{Huan-Huan Xie}
\author[mymainaddress]{Lei Li}
\author[mymainaddress]{Gang Lei}
\author[mymainaddress]{Ke Wang}
\author[mymainaddress]{Jian-Bo Deng}

\author[mymainaddress]{Xian-Ru Hu\corref{mycorrespondingauthor}}
\cortext[mycorrespondingauthor]{Corresponding author}
\ead{huxianru@lzu.edu.cn}

\address[mymainaddress]{School of Physical Science and Technology, Lanzhou University,
 Lanzhou 730000, People's Republic of China}

\begin{abstract}
Employing $ab$ $intito$ electronic structure calculations, we have investigated electronic and magnetic properties of the Zr-based quanternary Heusler alloys: ZrCoVIn, ZrFeVGe, ZrCoFeP, ZrCoCrBe and ZrFeCrZ (Z=In and Ga). Our $ab$ $intito$ calculation results show that all the alloys are (or nearly) spin-gapless semiconductors. All the alloys have large band gaps,  indicating the stability of them at room temperature.  The Slater-Pauling behaviours of these alloys are discussed as well. The values of Curie temperature of all the alloys are estimated. And it is found that the values of the Curie temperature for all our calculated quanternary Heusler alloys are higher than that of room temperature.
\end{abstract}


\end{frontmatter}


\section{Introduction}
Efficient spin injection from a ferromagnet to a semiconductor is very meaningful for the development of the performance of spintronic devices \cite{1}. Since the prediction of HM ferromagnetism of Heusler alloy NiMnSb by $ab$ $intito$ calculations in 1983 \cite{2}, half metallic ferromagnetism  (HMF) has attracted great interest \cite{19-2,19-3,19-5,19-6}. And the half metallic ferromagnetic Heusler alloys are good candidates for the application of spintronic devices, as these alloys often have high spin polarization, high Curie temperature compatible lattice structure. The improvement of computer science makes it possible to design materials on computers, which is efficient and less cost \cite{3-1,3-2,3-3}. Especially, computational material science enables the study of the new materials \cite{3-2,3-3}, which is useful to the experiments and practical applications.  

Recent studies have shown that there is a new class of materials, namely, spin-gapless semiconductors (SGS) \cite{4-1,4-2,4-3}. SGS was first predicted in diluted magnetic semiconductors PbPdO$_2$ by $ab$ $intito$ calculations \cite{4-1}. The excited carriers can be 100\% spin polarized with tunable capabilities, and the SGS may be more practical in the use of spintronic applications than half metals (HM). But there is a drawback that the Curie temperature (T$_C$) of this material is almost 180 K \cite{5-1,5-2}. As is known the Heusler alloys often have high T$_C$, the Heusler alloys may be a realization of SGS. And the Heusler compound Mn$_2$CoAl was first predicted to be a SGS with a high T$_C$ of 720 K \cite{4-2}. So the Heusler alloys are promising candidates for the future SGS use. 

The Heuler alloys consist of Zr element have attracted great interest. There are many theoretical and experimental studies on the semi-Heusler alloy ZrNiSn recently\cite{17-1,17-2,17-3,17-4}. Some investigations have shown that the Heusler alloys     Co$_2$ZrSn\cite{18-1,18-1-2}, Ni$_2$ZrSn\cite{18-1}, Ni$_2$ZrAl\cite{18-2}, Co$_2$ZrGe\cite{18-3} and ZrCoSb\cite{18-4} are half metals. And many other interesting properties have been found, for example, the alloy ZrPd$_2$Al is found to be superconductive\cite{16}.

In recent reports\cite{6-1,6-2,6-5,6-6,6-7,6-8}, some so-called LiMgPdSn or Y-type structure Heusler alloys with a formula of X$_{1}$X$_{2}$YZ have been discovered to be HMs by density function theory (DFT). Among the new structure Heusler alloys, MnCrVAl, MnCrTiSi, CoFeCrAl, CoFeTiAs, CoMnCrSi, MnVTiAs, FeVTiAs, FeCrTiAl, CrVTiAl, CoVTiAl, CoFeMnSi, CoFeCrAl and CoFeVSi are (or nearly) SGSs \cite{7-1,7-2}. As reported in \textbf{Refs.} \cite{4-2,8-1,8-2,8-3}, the ternary Heusler alloys Ti$_{2}$MnZ (Z=Al, Ga and In), Mn$_{2}$CoAl, Ti$_{2}$CoSi and Ti$_{2}$VAs are (or nearly) also SGSs. 

The Zr-Ti coupling quanternary Heusler alloys, ZrFeTiZ (Z=Al, Si and Ge) and ZrNiTiAl have been reported to be HMF very recently \cite{9}. This is the first prediction of HMF in the 4d-3d transition metal elements coupling quanternary Heusler alloys. As reported, these alloys have large half metallic band gaps and spin flip gaps. But they are just normal HMFs not SGSs.

It is very interesting that the Zr-Ti coupling  quanternary Heusler alloys have large HM band gaps and spin flip gaps, which means these alloys may be stable at room temperature. And there may be other Zr-based quanternary Heusler alloys with large HM band gaps, some probably being SGSs. Motivated by the above, we have designed the new Zr-based quanternary Heusler alloys: ZrCoVIn, ZrFeVGe, ZrCoFeP, ZrCoCrBe and ZrFeCrZ (Z=In and Ga). The electronic and magnetic properties of the alloys are investigated by $ab$ $intito$ calculations. ZrCoVIn, ZrCoCrBe, ZrCoFeP and ZrFeCrIn also have large band gaps of 0.98 eV, 0.71 eV, 0.41 eV and 0.80 eV. ZrFeCrGa and ZrFeVGe are semiconductors. The calculation results show that ZrCoVIn, ZrCoCrBe, ZrFeVGe, ZrCoFeP, ZrFeCrIn and ZrFeCrGa are (or nearly) SGSs. The Slater-Pauling behaviours of these alloys are discussed in detail. The values of Curie temperature of all the alloys are estimated. It is found that the values of the Curie temperature for all our calculated quanternary Heusler alloys are higher than that of room temperature. This means that all the calculated quanternary Heusler alloys keep being SGSs at room temperature and they may be practical in the future SGS use.

\section{Methods and details}
\label{sec:1}
The lattice optimization, electronic density of states (DOS), magnetic moment and band structure of the new Zr-based quanternary Heusler alloys are calculated by employing $ab$ $intito$ method. All our $ab$ $intito$ calculations are performed by using the full-potential local-orbital minimum-basis band structure scheme (FPLO) \cite{10-1,10-2} with generalized gradient approximation (GGA)\cite{11-1,11-2,11-3}. For the irreducible Brillouin zone, we use the $k$ meshes of 20$\times$20$\times$20 for all the calculations. The convergence criteria of self-consistent iterations is set to $10^{-6}$ to the density and $10^{-8}$ Hartree to the total energy per formula unit. 
 
As described in Refs. \cite{6-1,6-2}, the quanternary Heusler alloys has a so-called LiMgPdSn or Y-type structure (space group No.216, F$\bar{4}$3m). So all our calculations are performed in this lattice structure.

\section{Results and Discussions}
\label{sec:1}
In general, the quanternary Heusler alloys have a formula of $X_2YX_1Z$. In our calculations, the X$_2$ atom is Zr atom and Z is a main group element atom. X$_1$ and Y are the 3d transition metal element atoms,   which the atomic number X$_1$ is smaller than that of Y. According to group theory, the LiMgPdSn  or Y-type structure quanternary Heusler alloys have four Wychoff positions: 4a (0, 0 ,0), 4b ($\frac{1}{2}$, $\frac{1}{2}$, $\frac{1}{2}$), 4c ($\frac{1}{4}$, $\frac{1}{4}$, $\frac{1}{4}$) and 4d ($\frac{3}{4}$, $\frac{3}{4}$, $\frac{3}{4}$). The lattice structure is face centered cubic (FCC). In principle, the X$_1$, X$_2$, Y and Z atoms can occupy one of the 4a, 4b, 4c and 4d positions. By interchanging the positions of atoms in LiMgPdSn  or Y-type structure quanternary Heusler alloys, three possible types of atom arrangement are formed: Y-type ($\uppercase\expandafter{\romannumeral1}$), Y-type ($\uppercase\expandafter{\romannumeral2}$) and Y-type ($\uppercase\expandafter{\romannumeral3}$). $Zr$, Y, X$_1$ and Z atoms are arranged at different positions Y-type ($\uppercase\expandafter{\romannumeral1}$)=(4a, 4b, 4c, 4d), Y-type ($\uppercase\expandafter{\romannumeral2}$)=(4a, 4b, 4c, 4d) and Y-type ($\uppercase\expandafter{\romannumeral3}$)=(4a, 4b, 4d, 4c). In order to get the equilibrium structures of the Zr-based quanternary Heusler alloys, the geometry optimization are firstly performed in their three different configurations by calculating the total energies as a function of lattice constants. From the calculated results of total energies at equilibrium lattice constants, we find that Y-type ($\uppercase\expandafter{\romannumeral1}$) is the most stable one of the three structures for both spin-polarization (FM phase) and non-spin-polarization (NM phase). And FM phase is more stable than NM phase in Y-type ($\uppercase\expandafter{\romannumeral1}$).This is in agreement with the previous papers \cite{6-1,6-2,6-5,6-6,9}. The obtained equilibrium of lattice results in Y-type ($\uppercase\expandafter{\romannumeral1}$) are presented in \textbf{Table 1} for the FM phases. So we will only discuss the quanternary Heusler alloys in Y-type ($\uppercase\expandafter{\romannumeral1}$) structure for the FM phases. 
\subsection {Slater-Pauling behaviours}
\label{sec:1-1}
\textbf{Table 2} shows the magnetic moment of all our calculated alloys under the equilibrium lattice constants.

All the calculated quaternary Heusler alloys  have very large band gaps. It means they may keep their magnetic properties at room temperature. 

We will have a brief discussion of the Slater-Pauling behaviours based on the values of total magnetic moment in \textbf{Table 2}. The calculated total magnetic moment per formula unit, 3 $\mu_B$, for the quanternary Heusler alloys ZrCoVIn, ZrCoCrBe, ZrFeCrIn, ZrFeCrGa and ZrFeVGe, obeys the Slater-Pauling behaviour which can be expressed by,
\begin{equation}
M_{tot}=(Z_{tot}-18) \mu_{B}
\end{equation}
here $M_{tot}$ and $Z_{tot}$ are the total magnetic moment per formula unit and the number of total valence electrons in each the above alloys. $Z_{tot}$ is 21.  The values of magnetic moment per formula unit is 2 $\mu_B$ for the calculated quanternary Heusler alloys ZrCoFeP. The Slater-Pauling behaviour which is obeyed by ZrCoFeP can be expressed by,
\begin{equation}
M_{tot}=(Z_{tot}-24) \mu_{B}
\end{equation}
here $M_{tot}$ and $Z_{tot}$ have the same meaning as previous. The value of $Z_{tot}$ is 26.

The investigations of the Slater-Pauling behavours of usual full and inverse Heusler alloys can be found in Refs. \cite{12-1,12-2}. Similar to the discussions in Refs. \cite{12-1,12-2}, we present the possible hybridizations between miority - spin $d$ orbitals at different occupations in the case of 21 and 26 valence electrons in \textbf{Figure 1} and \textbf{Figure 2}. As described in \textbf{Ref \cite{12-1} }for both figures, d$_{1,...,5}$ orbitals correspond to the d$_{xy}$, d$_{yz}$, d$_{zx}$, d$_{3z^2-r^2}$ and d$_{x^2-y^2}$ orbitals, respectively. 

\textbf{Figure 1} is in corresponding to the alloys ZrCoVIn, ZrCoCrBe, ZrFeCrIn, ZrFeCrGa and ZrFeVGe. For these alloys, the 3d transition metallic elements X$_1$ and Y sit at the sites with the same symmetry, so their 3d orbitals hybridize similar to which described in \textbf{Refs. \cite{12-1,12-2}}. The 3d orbitals of X$_1$ and Y atoms hybride, creating five 3d bonding and nonbonding states. And the five X$_1$-Y bonding states hybride with the 4d orbitals of the Zr atom, creating again bonding and antibonding states. Similar to the Sc$_2$ and Ti$_2$ based inverse Heusler compounds\cite{12-2}, the triple-degeneration $t_{1u}$ states and the double-degeneration $e_u$ states are in very high energy level. So both the $t_{1u}$ and $e_u$ states are empty. And the spin-down gap is created between the non-bonding $t_{1u}$ states and the bonding $t_{2g}$ state. As can be seen in \textbf{Figure 1}, one bonding $t_{2g}$ state and $e_g$ state are below the Fermi level in the hybridization schematic. The un-shown 1$\times$s state and 3$\times$p state are also below the Fermi level. In total there are 9 states below Fermi level. The total magnetic moment $M_{tot}$ (in $\mu_{B}$) is just the difference between the number of occupied spin-up states and occupied spin-down states. We can directly deduce the number of occupied spin-up states: $N\uparrow=Z_{tot}-N\downarrow$. So the total magnetic moment: $M_{tot}=(N\uparrow-N\downarrow) \mu_{B}=(Z_{tot}-2\times N\downarrow)\mu_{B}=(Z_{tot}-18) \mu_{B}$. So the Heusler alloys ZrCoVIn, ZrCoCrBe, ZrFeCrIn, ZrFeCrGa and ZrFeVGe obey the Slater-Pauling behaviour expressed by \textbf{Equation(1)}.

As for ZrCoFeP, we should focus on the hybridization schematic in \textbf{Figure 2}. $X_1$, Y and Z are Fe, Co and P, respectively. This case is similar to the full-Heusler alloys.\cite{12-1} As can be seen in \textbf{Figure 2}, the $t_{1u}$ state is below Fermi level and $e_u$ state is above Fermi level. The the spin-down gap is created between the nonbonding $t_{1u}$ and $e_u$ states. From \textbf{Figure 2}, we can get that $t_{2g}$ and $e_g$ states are below Fermi level. Considering the 1$\times$s and 3$\times$p states, there are 12 states below Fermi level in total. We can directly deduce the total magnetic moment:  $M_{tot}=(N\uparrow-N\downarrow) \mu_{B}=(Z_{tot}-2\times N\downarrow)\mu_{B}=(Z_{tot}-24) \mu_{B}$. That is why ZrCoFeP obeys the Slater-Pauling behaviour expressed by \textbf{Equation(2)}.

It is very interesting that the number of the valence electrons for our calculated spin-gapless semiconductors (SGS) is either 21 or 26, which is similar to the results described in \textbf{Ref \cite{7-1}}. And the Slater-Pauling behaviours obeyed by our calculated SGSs are also similar to the discussions in \textbf{Ref \cite{7-1}}.
\subsection {Electronic structure properties}
\label{sec:1-2}
In this section, we will discuss the properties of spin-gapless semiconductors (SGS). \textbf{Figure 3} shows the total density of states of all the calculated alloys. And \textbf{Figure 4} shows the partial density of states (PDOS) of ZrCoVIn under the equilibrium lattice constants of all our calculated quanternary Heusler alloys.

Firstly, we focus on the total DOS. In each of  the cases, there is a large gap in the spin down band structure and the Fermi level falls within this gap. In the spin up band structure, for ZrCoVIn and ZrFeCrIn, the valence and conduction bands touch each other and the Fermi level falls within a zero-energy gap, which forms a valley in the spin up band structure. As described in \textbf{Refs.} \cite{4-1}, ZrCoVIn and ZrFeCrIn can be classified as SGSs. So no energy is required to excite electrons from the valence band to the conduction band, which is the same phenomenon that can be seen for the Hg-based gapless semiconductors and graphene \cite{4-1}. And more interesting is that for an excitation energy up to the band gap energy of the spin channel and the holes can also be 100\% spin polarized. So the carriers two can be possible fully polarized in the two SGSs. Therefore they can be used as spintronic materials with superior performance to half metals and diluted magnetic semiconductors \cite{4-1,4-2}. In the cases of ZrCoCrBe and ZrCoFeP, in the spin up band structure, there is a small overlap  of the band being located and above and below the Fermi level although no band-crossing occurs. So the quanternary Heusler alloys ZrCoCrBe and ZrCoFeP are almost SGSs. In the case of ZrFeCrGa, it is clear to see there is a large band gap in the spin-down channel. And a close look at the band structure reveals that there is a very narrow band gap of 0.02 eV. In the spin up band structure, the valence and conduction bands touch each other but the Fermi level falls within a narrow band gap. So ZrFeCrGa is very close to a SGS. In the case of ZrFeVGe, there is a large band gap in both spin down and up channels but the two gaps are not located at the same energy region. And a close look at the band structure reveals that there is a band gap below the Fermi level in the spin up band which touches the Fermi level resulting in an almost vanishing DOS below the Fermi level. So the Fermi level slightly crosses the spin-down conduction band and the spin-up valence band. As a result, the quanternary Heusler alloy ZrFeVGe can be classified as an indirect spin-gapless semiconductor.

Next the partial density of states (PDOS) will be discussed. From \textbf{Figure 4}, we can get that for the quanternary Heusler alloy ZrCoVIn, the 3d states of V and Co atoms make the most contributions to the total DOS near the Fermi levels. The 4d states of Zr make the most contributions to the total DOS of the Zr states near the Fermi levels. And the 5p states of In states make the most contributions to the total DOS near the Fermi levels of the Z states. From the PDOS, it can be seen there are hybridizations between V-3d, Zr-4d and Co-3d states around the Fermi levels. We can get similar conclusions for the other calculated alloys.
\subsection {Curie temperature}
\label{sec:1-3}
As is known, the Curie temperature, $T_C$, of the magnetic materials is crucial for the practical applications. So in this section, we would comment on the expected Curie temperature. 

The previous investigations on multi-sublattice half-metallic Heusler compounds have shown that Curie temperature is more or less proportional to total spin magnetic moment (or the sum of the absolute values of the atomic spin magnetic moments in the case of ferrimagnets) since Curie temperature is mainly determined by the nearest neighbor inter-sublattice exchange interactions  \cite{13-1,13-2,13-3,13-4,13-5,13-6}. As described \textbf{in Refs.} \cite{4-2,8-1}, it is found experimentally that the $T_C$ of Mn$_2$CoAl is 720 K and the the sum of the absolute values of the spin moments is 5.47 $\mu_B$. Based on the empirical value and according to \textbf{Table 2}, we can estimate that the value of $T_C$ for the quanternary Heusler alloy ZrCoFeP is 320 K. As the sum of the absolute values of the atomic spin magnetic moments for ZrCoFeP is the lowest of all our calculated quanternary Heusler alloys, the values of $T_C$ for all our calculated quanternary Heusler alloys are higher than that of room temperature. So the SGSs may probably be stable at room temperature. And they may be candidates for the future spin-gapless semiconductors applications.
\section{Conclusions}
\label{sec:2}
In conclusion, we have investigated some Zr-based quanternary Heusler alloys by employing  $ab$ $intito$ calculations. It is found that the Zr-based quanternary Heusler alloys ZrCoVIn, ZrCoCrBe, ZrFeVGe, ZrCoFeP, ZrFeCrIn and ZrFeCrGa are (or nearly) SGSs with large band gaps by studying the DOS. The Slater-Pauling behaviours of these alloys are discussed as well. The Curie temperature for these alloys have also been estimated, and the results show that the values of the Curie temperature for these alloys are higher than that of room temperature. So these alloys can be the potential candidates for the future SGS applications.




\newpage
 \bibliographystyle{elsarticle-num}
 \bibliography{Reference.bib}

\clearpage
\begin{table}
\caption{The results of the lattice optimization of all our calculated quanternary Heusler alloys in Y($\uppercase\expandafter{\romannumeral1}$) structure for FM phase.}{\label{Table 1}}
\begin{tabular}{ccccccccccccccccccc}
\hline
ZrYX$_1$Z & a$_{opt}$ (\AA) & E$_{tot}$ (Ry)&band gap (eV)& \\
\hline
    
	ZrCoVIn& 6.468 & -23650.670779&0.98& \\
    ZrCoCrBe & 6.013&-12116.709912&0.71&\\
	ZrFeCrIn & 6.419&-23612.431029&0.80&\\
	ZrFeCrGa & 6.184&-15734.120866& $\uparrow$0.02 $\downarrow$0.71
	&\\
	ZrFeVGe&6.210&-15840.940640& $\uparrow$0.41 $\downarrow$0.81& \\
	ZrCoFeP&5.944&-13215.295374&0.41&\\
\hline
\end{tabular}
\end{table}
\clearpage
\begin{table}
\caption{The partial and total magnetic moments of all our calculated Heusler alloys in type ($\uppercase\expandafter{\romannumeral1}$) structure under the equilibrium lattice constant. M $_{abs}$ is the sum of magnetic moments of all the atoms for each alloy.}{\label{Table 2}}
\begin{tabular}{cccccccccccccccccc}
\hline
ZrYX$_1$Z &M$_{X_1}$ ($\mu_B$)
	
	 & M$_{Zr}$ ($\mu_B$) & M$_Y$ ($\mu_B$) & M$_Z$ ($\mu_B$) & $M_{tot}$ ($\mu_B$)&M$_{abs}$ ($\mu_B$)\\
\hline
	ZrCoVIn & 2.89 & 0.19 & 0.04 & -0.12 & 3.00&3.25\\
    ZrCoCrBe & 3.21 & -0.17 & 0.14 & -0.18 &3.00&3.72\\
	ZrFeCrIn & 3.45 & -0.27 & -0.03 & -0.15 &3.00&3.91 \\
	ZrFeCrGa&3.12 & -0.29 & 0.32 & -0.15 &3.00&3.88\\
	ZrFeVGe&2.58&0.00& 0.57&-0.15&3.00&3.32  \\
	ZrCoFeP&1.19&-0.24&0.99&0.06&2.00&2.49   \\
	\hline
\end{tabular}
\end{table}

\newpage
{\bf Figure captions}

\textbf{Figure 1:} Possible hybridizations between spin-down orbitals siting at different sites in the case of 21 valence electrons for our calculated quaternary Heusler  alloys.

\textbf{Figure 2:} Possible hybridizations between spin-down orbitals siting at different sites in the case of ZrCoFeP.

\textbf{Figure 3:} The total density of states (DOS) for all our calculated quaternary Heusler  alloys.

\textbf{Figure 4:} The partial density of states (PDOS) for ZrCoVIn.

\clearpage
\textbf{Figures}
\begin{figure}
  \centering
  \includegraphics[width=1\textwidth]{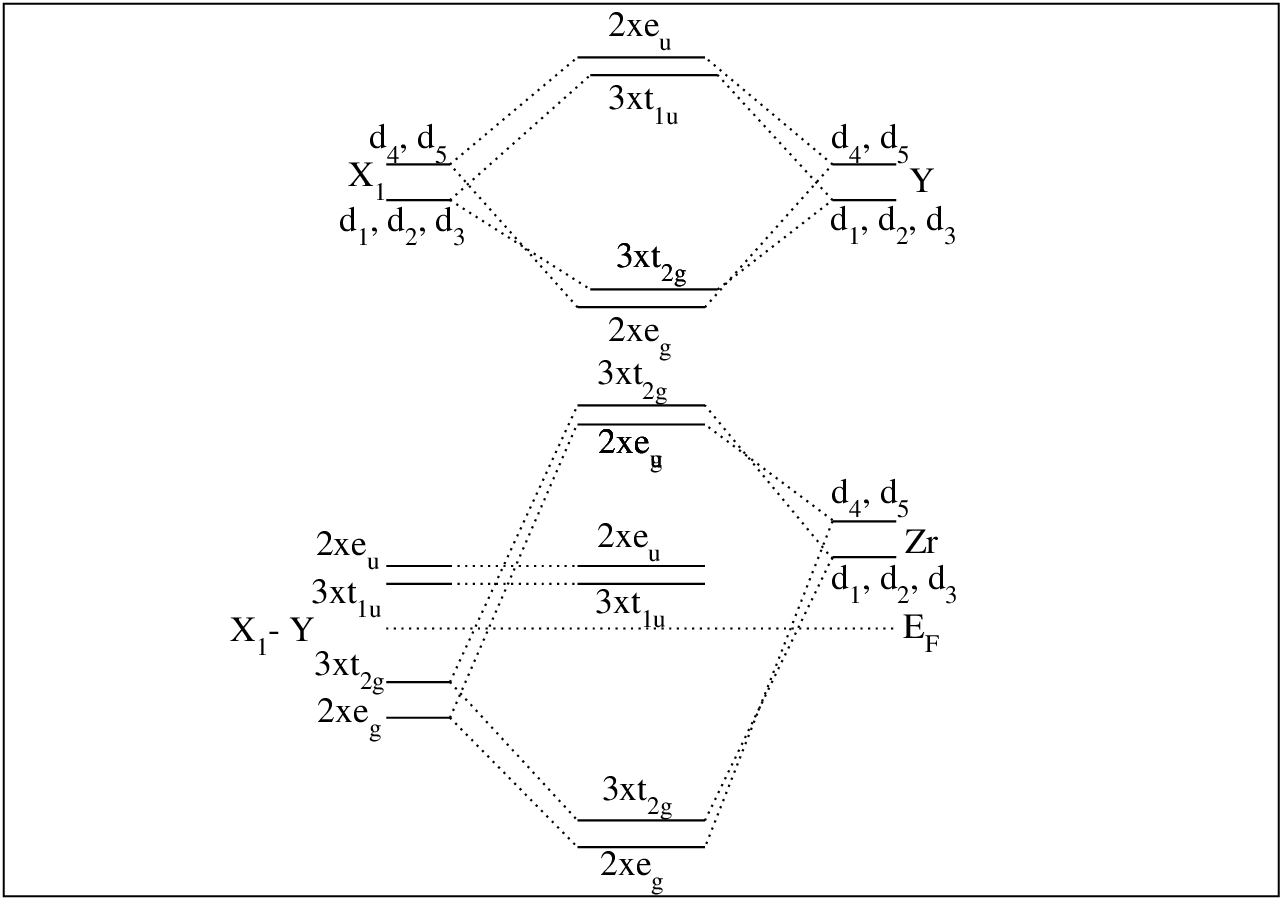}
\caption{ }
\label{Figure1}       
\end{figure}
\clearpage
\begin{figure}
  \centering
  \includegraphics[width=1\textwidth]{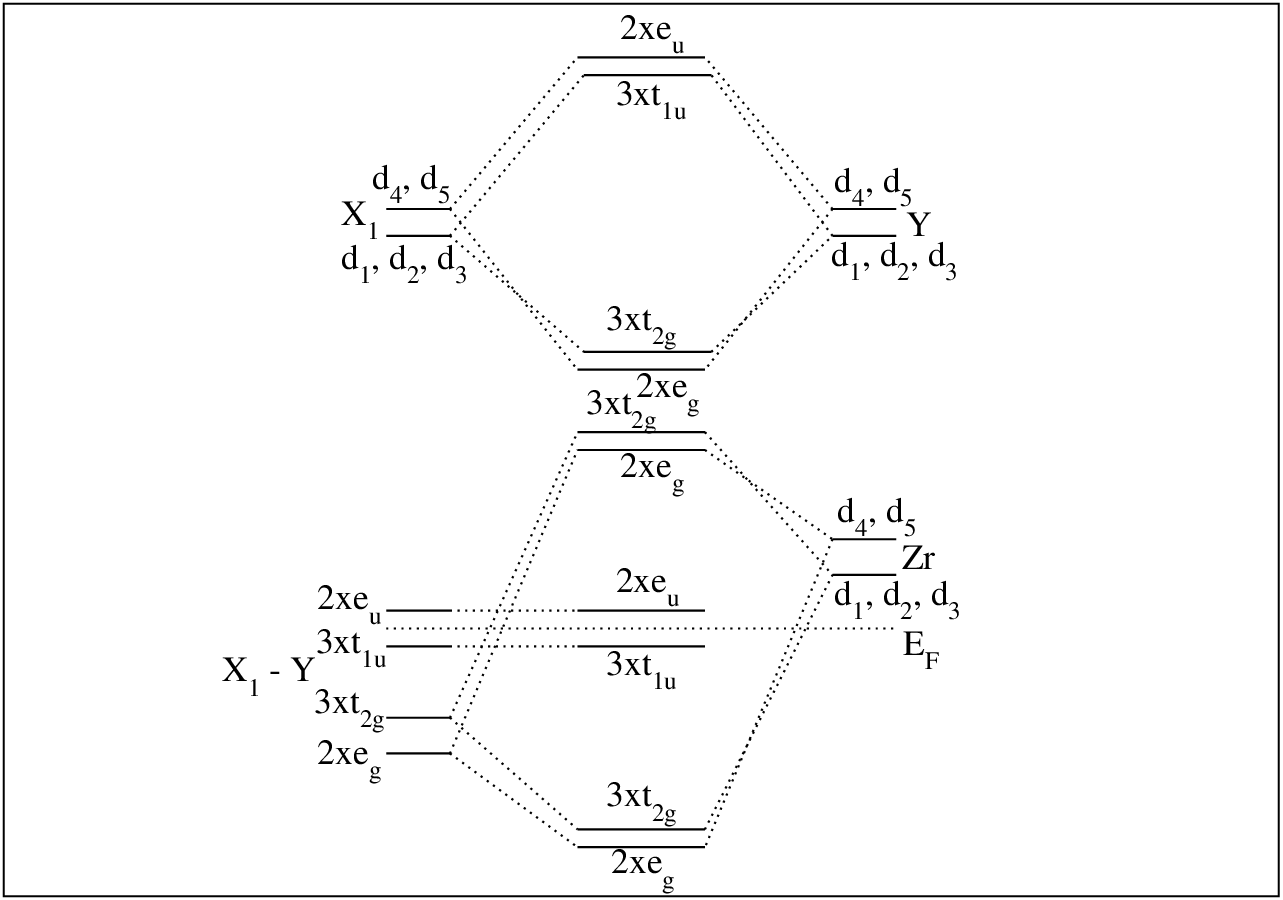}
\caption{ }
\label{Figure2}       
\end{figure}

\clearpage
\begin{figure}
  \centering

  \includegraphics[width=1\textwidth]{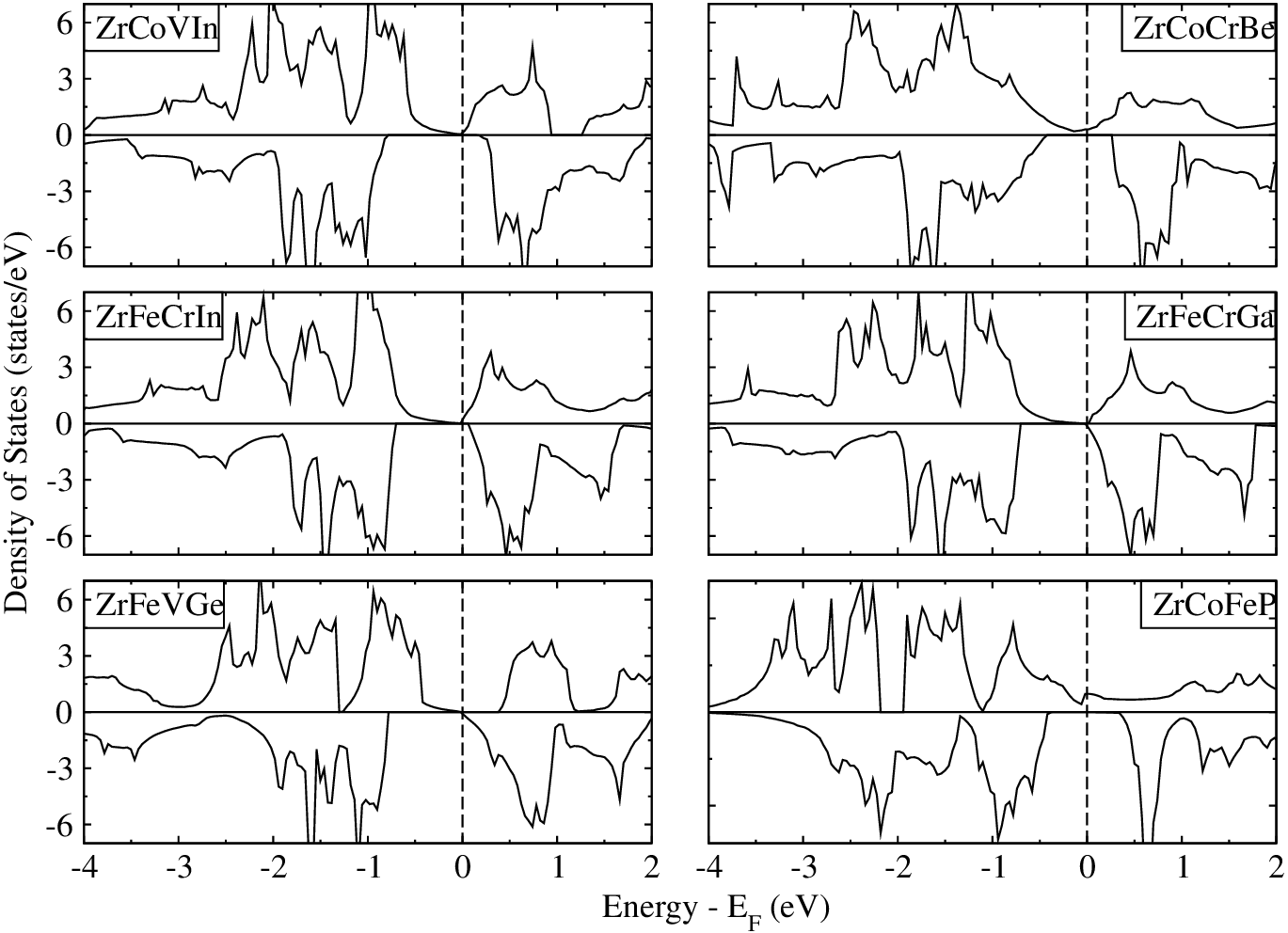}
\caption{ }
\label{Figure3}       
\end{figure}
\clearpage
\begin{figure}
  \centering
  \includegraphics[width=1\textwidth]{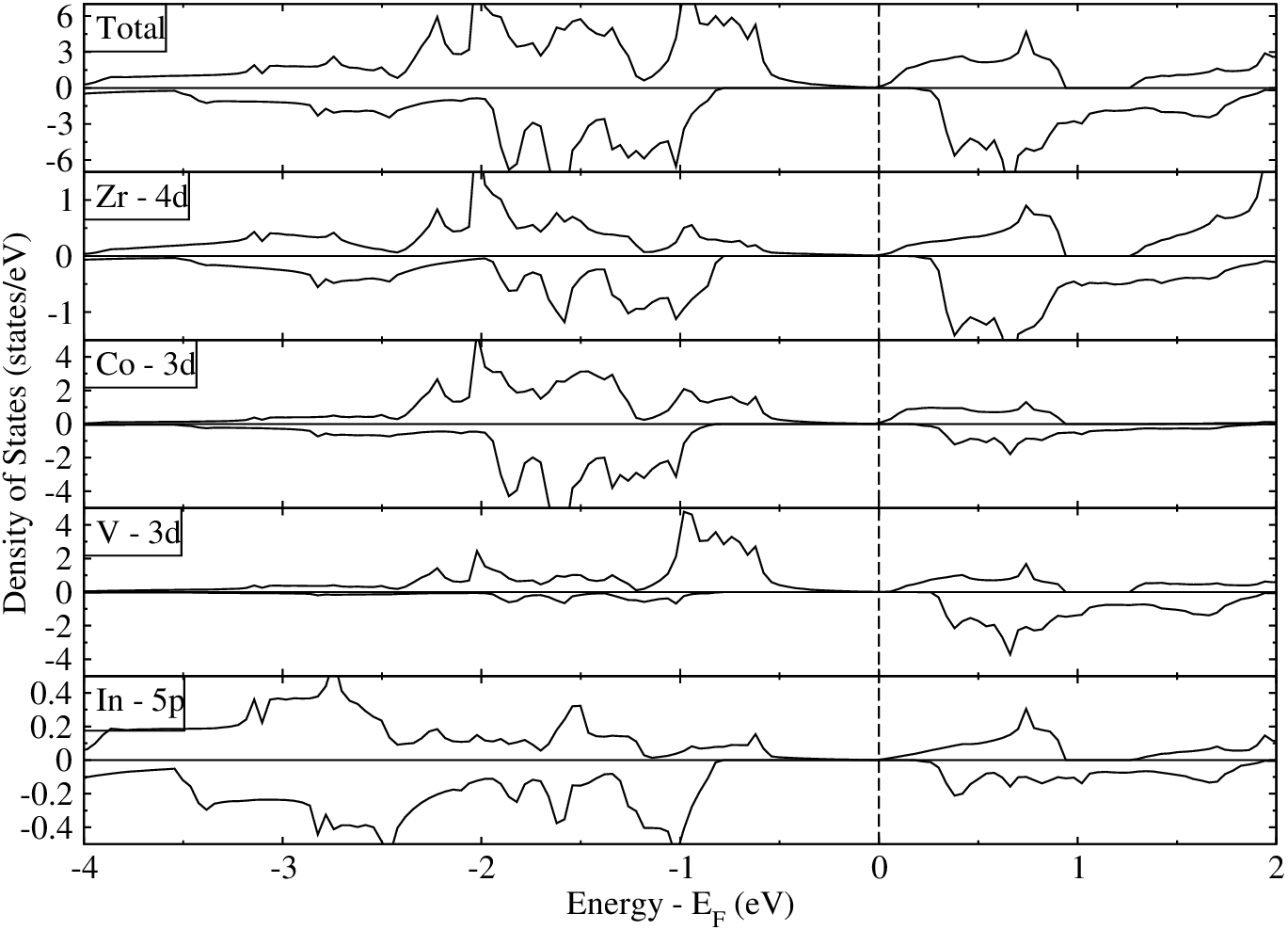}
\caption{ }
\label{Figure4}       
\end{figure}

\end{spacing}{2.0}
\end{document}